\pdfoutput=1 
\documentclass[sigconf,nonacm,10pt]{acmart}
 
\setcopyright{none}
\usepackage{tabularx}
\usepackage{bm}
 
\newcolumntype{L}{>{\raggedright\arraybackslash}X}
 
\begin{document}
 
 
\title[Self-Expansion Through AI]{Self-Expansion Through AI: Latent Profile Analysis of AI Companionship and Wellbeing Outcomes}
 
 
\author{Ali Sahandi}
\orcid{0009-0004-0232-0072}
\email{sahandiali.sci@gmail.com}
\affiliation{%
  \institution{Independent Researcher}
  \city{Syracuse}
  \state{NY}
  \country{USA}
}
 
\author{Jaime Banks}
\orcid{0000-0002-7598-4337}
\email{banks@syr.edu}
\affiliation{%
  \department{School of Information Studies}
  \institution{Syracuse University}
  \city{Syracuse}
  \state{NY}
  \country{USA}
}
 
\renewcommand{\shortauthors}{Sahandi and Banks}
 
 
\begin{abstract}
People increasingly form meaningful relationships with conversational AI. Current variable-centered literature offers a mixed picture of relational patterns and how they associate with wellbeing. Taking a person-centered approach, we applied latent profile analysis to secondary data from 467 active conversational AI users, evaluating five indicators of machine companionship: utilitarian motivation, eudaimonic exchange, connective coordination, mind perception, and inclusion of other in the self. Five profiles emerged, ordering parsimoniously from objectified-utilitarian to intimate-bonded. Conceptually divergent utilitarian motivation and companionate engagement co-occurred, rather than diverging. Furthermore, deep companionship was achieved without users attributing a mind to the AI. Profile membership was strongly associated with relatedness need satisfaction, tracking upward across the groups, but was only weakly associated with loneliness, indicating AI fulfills immediate relational needs but may not resolve systemic human disconnection. Ultimately, these findings characterize human-AI relationships as graded configurations spanning a superordinate continuum of self-expansion.
\end{abstract}
 
 
 
\maketitle

\section{Introduction}
Artificial intelligence is increasingly entering everyday life not only as a tool for information, assistance, and task support, but also as a social presence designed to converse, remember, encourage, and remain available over time~\cite{rogge_defining_2023, strohmann_toward_2023}. Across companion chatbots~\cite{skjuve_my_2021}, voice assistants~\cite{schweitzer_servant_2019}, and more recent conversational and generative AI systems~\cite{chaturvedi_social_2023, pentina_consumermachine_2023}, this shift has positioned human--AI relationships as an increasingly important topic within human--computer interaction research, as the significance of these systems depends not only on their functional capabilities but also on the forms of social connection and relational meaning users perceive with them~\cite{chaturvedi_social_2023, pentina_consumermachine_2023, rogge_defining_2023}.

Within this broad context, machine companionship (MC) has emerged as a construct for capturing a positively experienced, enduring, and socially meaningful connection with AI that goes beyond purely instrumental use~\cite{banks_measuring_2026, banks_conceptualization_2026}. Recent work suggests that such human--AI relationships can take multiple forms, ranging from largely functional engagement to more companionate, friend-like, romantic, or deeply integrated ties~\cite{alabed_more_2024, ebner_understanding_2026, schweitzer_servant_2019, tschopp_servant_2023}. Understanding these forms matters in part because different patterns of human--AI relationship may shape psychosocial outcomes such as loneliness and relatedness need satisfaction (RNS; i.e., the extent to which a relationship provides a sense of care, closeness, and belonging), both of which are central outcomes in broader theories of human connection and wellbeing~\cite{baumeister_need_1995, ryan_self-determination_2000}.

Existing literature has advanced the conceptualization of MC, identifying recurring relational templates~\cite{alabed_more_2024, banks_measuring_2026, schweitzer_servant_2019, tschopp_servant_2023} and examining how these templates relate to psychosocial outcomes~\cite{banks_measuring_2026, liu_heterogeneous_2025}. However, attempts to understand user heterogeneity remain limited in two ways. First, extant work relies largely on variable-centered approaches, which are valuable for identifying general patterns of association but treat the sample as relatively homogeneous, leaving open how multiple relational tendencies co-occur within the same person~\cite{laursen_person-centered_2006}. A person-centered approach addresses this gap by identifying subgroups of individuals who share a common configuration across several attributes~\cite{bergman_person-oriented_1997}. Yet, a second limitation persists: while recent person-centered studies have identified distinct user typologies, they largely segment users based on narrow variables, such as assigned roles, personality traits, or functional motives~\cite{schweitzer_servant_2019, wienrich_long-term_2023}. What remains missing is a person-centered approach grounded in the multidimensional relational experience itself.

To address this gap, the present study conducts secondary analysis of data from a variable-centered study ~\cite{banks_measuring_2026}, instead using Latent Profile Analysis (LPA ~\cite{spurk_latent_2020}) to identify distinct patterns of MC. From that sample of users engaging conversational AI for social purposes ($N = 467$), we profile users across three layers of engagement: motivational orientation (utilitarian motivation), social-cognitive construal (mind perception, inclusion of other in the self), and experiential qualities (eudaimonic exchange, connective coordination). We then evaluate how these resulting profiles associate with loneliness and RNS outcomes.

Our analysis identified five distinct profiles that form an underlying continuum, ranging from objectified-utilitarian to intimate-bonded. We interpret this continuum as representing a gradient of self-expansion~\cite{aron_self-expansion_2022}, reflecting how users leverage the AI to augment their own resources, perspectives, and identities. Viewed through this theoretical lens, functional utility and social intimacy operate together rather than as competing motives. We demonstrate that achieving deep companionship and self-expansion does not require the user to perceive the AI as a minded agent. Finally, while profile membership was strongly associated with RNS, which rose steadily across the continuum, it showed only a weak relationship with general loneliness.

\section{Background and Related Work}

\subsection{Conceptualizing Human--Machine Companionship}

Research on human relations with artificial agents spans multiple disciplinary traditions, and observations that people respond to machines as if socially significant date back at least to the 1960s, when users engaged the early conversational program ELIZA as though it understood them and confided personal concerns to it despite its simplicity~\cite{weizenbaum_elizacomputer_1966, weizenbaum_computer_1976}. This tendency was established experimentally from the 1990s in the Media Equation and the Computers Are Social Actors (CASA) paradigm, which showed that people automatically apply human social rules and expectations to computers in response to minimal social cues, even while consciously recognizing that computers are not human~\cite{nass_machines_2000, reeves_media_1996}. Studies of more overtly social technologies extended this line to affective bonds, showing that people can form emotionally significant attachments to sociable robots even while fully aware of their artificiality~\cite{turkle_relational_2006}. More recently, \citet{gambino_building_2020} extended CASA for contemporary AI, arguing that social responses are shaped not only by human--human scripts but also by human--media scripts learned through repeated interaction with increasingly social technologies. Moving beyond this paradigm, \citet{van_der_goot_disentangling_2023} distinguished a Media Evocation perspective, in which machines are treated not as stand-ins for humans but as evocative objects whose ambiguous, ``betwixt and between'' status~\cite[p.~18]{van_der_goot_disentangling_2023} prompts users to reflect on what the machine is, ontologically. Approaching communication as an agent-agnostic process, \citet{banks_toward_2020} hold that all communicative engagement, including interactions with machines, is grounded in a fundamentally social process of encoding and decoding. Consequently, relational bonds with AI should not be viewed as simulative errors or parasocial mimicry, but rather as operationally actual meaning-making and valid subjective experiences~\cite{banks_ghosting_2026, banks_measuring_2026}.

The extent to which individuals engage with artificial agents as social partners has often been explained in terms of anthropomorphism, defined as the attribution of humanlike mental states and intentions to nonhuman entities~\cite{epley_seeing_2007}. This tendency varies across individuals, with dispositions such as sociality motivation and need for affiliation shaping the degree to which people anthropomorphize machines~\cite{waytz_who_2010}. However, social engagement with robots is not limited to anthropomorphism. \citet{norskov_sociomorphing_2020} argue that users often engage in sociomorphing, which involves directly perceiving a robot's machine-native capacities for social coordination. While behaviors like programmed turn-taking mimic human patterns, they are executed as structural, algorithmic affordances. Thus, users can interact with a robot as an ``idiosyncratic social other''~\cite[p.~58]{norskov_sociomorphing_2020} without attributing human consciousness to the machine. Closely related to these processes is mind perception. \citet{gray_dimensions_2007} conceptualized mind perception, defined as the attribution of mental capacities to an entity, along two dimensions: agency, referring to the capacity for intention, planning, and action, and experience, referring to the capacity for sensation, emotion, and subjective feeling. \citet{koban_it_2024} found that perceiving a social robot as capable of emotion was the strongest predictor of explicitly ascribing it a mind, identifying affective capacity as a key basis on which people may come to treat an artificial agent as an authentically minded partner rather than a mere tool.

Perceiving the agent as capable of feeling and responding supports a reciprocal bond with a perceived partner, and attachment theory helps explain how such engagement can deepen into an emotionally significant tie~\cite{bowlby_attachment_1969}. First developed for infant--caregiver ties~\cite{ainsworth_patterns_1978} and later extended to adult relationships~\cite{hazan_romantic_1987}, the theory holds that a close other can serve as a safe haven in times of distress and a secure base from which to engage the world, with stable individual differences in attachment style shaping how readily such bonds form~\cite{mikulincer_attachment_2010}. Recent work extends this framework to human--AI relationships, showing that people can form attachment-like bonds with artificial agents and engage in proximity-seeking and safe-haven behaviors~\cite{gillath_attachment_2021, xie_attachment_2022}. Such bonds are most likely to intensify under distress, loneliness, or unmet companionship needs and to develop gradually through self-disclosure, trust, and routine interaction~\cite{pentina_exploring_2023, skjuve_my_2021, xie_attachment_2022}. In some cases, emotional dependence on an AI companion can resemble human relational dependency~\cite{laestadius_too_2024}, carrying risks of overreliance or displacement of human relationships~\cite{fang_how_2025}. Because attachment centrally concerns the satisfaction of needs for security and belonging~\cite{baumeister_need_1995, mikulincer_attachment_2010}, this framework links human--AI bonds directly to the relational and emotional outcomes. While this literature shows that people can form meaningful bonds with AI, it also highlights the need for a clearer conceptual definition of MC itself.

\citet[p.~13]{banks_conceptualization_2026} define machine companionship as ``an autotelic, coordinated connection between a human and machine that unfolds over time and is subjectively positive.'' Building on this definition, \citet{banks_measuring_2026} developed a dedicated multidimensional measure of MC. Although the item pool was grounded in five theoretical facets (dyadism, autotelicity, coordination, temporality, and positive valence), factor analyses in a sample of active AI companion users ($N = 467$) and a confirmatory sample ($N = 249$) indicated that MC is best captured by two dimensions that entangle those facets: (a) eudaimonic exchange: a purposeful and mutually elevating give-and-take, and (b) connective coordination: a sense of interpersonal synchronization and bonding. This operationalization treats MC as a relational process rather than a simple perception, a one-directional attachment, or a functional use pattern, and it provides the conceptual and measurement basis for the present study.

\subsection{Heterogeneity in Machine Companionship: A Person-Centered Approach}

Most research on machine companionship has been variable-centered, examining how features of the relationship relate to one another or to outcomes across a sample as a whole. Such work identifies general patterns but assumes they hold uniformly across users, and if the population is heterogeneous, associations estimated across the sample may describe no actual subgroup well~\cite{bergman_person-oriented_1997, laursen_person-centered_2006}.

Person-centered research addresses this limitation by grouping individuals rather than variables, and it indicates that people do not relate to machines uniformly but fall into distinct, recurring profiles. Across companion chatbots, voice assistants, social robots, and other smart devices, studies have identified typologies of between two and seven groups (Table~\ref{tab:typologies}). While useful, these typologies largely segment users on the basis of isolated variables, and categorizing a relationship by a designated role or functional motivation alone may serve as only a partial proxy for relational depth. What remains missing is a person-centered approach grounded in a multidimensional measure of the relational experience itself.

To address this gap, the present study uses LPA to examine how people relate to AI across multiple dimensions of engagement, profiling users across three layers: motivational orientation, social-cognitive construal, and experiential qualities. This leads to our first research question:

\noindent\textbf{RQ1:} What latent profiles can be identified in human experiences of companionship with AI?


\begin{table*}[t]
\caption{Person-centered typologies of human--machine companionship reported in prior work.}
\Description{A four-column table listing eight prior studies. For each study, the table gives the interaction context, the profiles or clusters that study identified, and the variables on which users were classified. Contexts range from conversational AI and companion chatbots to voice assistants, social robots, embodied agents, and smart devices. The number of identified groups ranges from two to seven, and classification bases include personality traits, assigned roles, usage patterns, anthropomorphism, and user agency.}
\label{tab:typologies}
\small
\begin{tabularx}{\linewidth}{@{}p{0.16\linewidth} p{0.15\linewidth} L p{0.21\linewidth}@{}}
\toprule
\textbf{Study} & \textbf{Context} & \textbf{Identified profiles / clusters} & \textbf{Basis of classification} \\
\midrule

Alabed et al.~\cite{alabed_more_2024} & Conversational AI & Functional; Aspiring; Committed; Replacement & Self-congruence and self--AI integration \\
\addlinespace

Liu et al.~\cite{liu_heterogeneous_2025} & Companion chatbots & Disengaged Light; Well-Adjusted Moderate; AI-Wary Light; Lonely Moderate; Fulfilled Dependent; Lonely Light; Socially Challenged Frequent & Usage patterns, loneliness, and problematic use \\
\addlinespace

Müller et al.~\cite{muller_chatbot_2019} & Chatbots (Alexa) & Introverted careless distrusting; Conscientious curious trusting; Careless dishonest trusting & HEXACO personality traits and trust \\
\addlinespace

Schweitzer et al.~\cite{schweitzer_servant_2019} & Voice assistants & Servant; Partner; Master & Assigned roles and anthropomorphism \\
\addlinespace

Smrke et al.~\cite{smrke_aging_2022} & Embodied agents & Low motivation; Selective (physical well-being); Selective (psychological well-being); High motivation & Importance of daily living activities \\
\addlinespace

Toivonen and Lelli~\cite{toivonen_varieties_2024} & Smart devices & Controller; Collaborator; Detached user; Victim & User agency versus device agency \\
\addlinespace

Wienrich et al.~\cite{wienrich_long-term_2023} & Voice assistants & Friend cluster; Non-friend cluster & Implicit friendship quality and usage behavior \\
\addlinespace

Yi and Park~\cite{yi_potential_2024} & Social robots & Burdensome Machine for Information; Trusted Friend; Emotionally Intelligent Device; Energizing Gadget & User expectations and anthropomorphism \\

\bottomrule
\end{tabularx}
\end{table*}

At the level of motivational orientation, Utilitarian Motivation captures the extent to which a person engages the AI for instrumental, task-oriented ends, valuing it for its usefulness, effectiveness, and efficiency rather than for the experience of the interaction itself, a functional orientation that consumer research contrasts with hedonic, intrinsically valued engagement~\cite{voss_measuring_2003}. It is included because it addresses whether the AI is engaged primarily as a functional resource or as a relational partner valued for its own sake. Prior work suggests that human--AI companionship often contains both instrumental and relational elements, with meaningful variation across users lying in the relative balance between these orientations rather than in either alone~\cite{banks_measuring_2026, banks_conceptualization_2026, schweitzer_servant_2019}.

At the level of social-cognitive construal, Mind Perception and Inclusion of Other in the Self capture how the AI is cognitively represented as a social entity. Following \citet{gray_dimensions_2007}, Mind Perception refers to the attribution of a mind along two dimensions: agency, the capacity to plan and act, and experience, the capacity to feel and be affected. Such perceptions are consequential because they increase social responding and make it more likely that the AI is construed as a genuine social other rather than a mere tool~\cite{banks_conceptualization_2026, hacker_chatbots_2025}. Inclusion of Other in the Self captures perceived closeness and self--other overlap, indicating the extent to which the AI becomes incorporated into the user's sense of self~\cite{aron_inclusion_1992}. Rooted in self-expansion theory, this cognitive overlap reflects a fundamental human drive to increase one's own capabilities and perspectives by integrating those of a partner~\cite{aron_self-expansion_2022}. In the context of MC, this expansion process facilitates both relational closeness and the anticipation of functional and social rewards ~\cite{banks_measuring_2026}.

At the level of experiential qualities, Eudaimonic Exchange and Connective Coordination are included because they capture the two core relational dimensions of MC as an ongoing relational process~\cite{banks_measuring_2026}. Eudaimonic Exchange reflects whether interacting with the AI feels enriching, growth-promoting, and personally significant, which helps distinguish deeper forms of companionship from casual or entertainment-based engagement. Connective Coordination captures the sense of mutual involvement, attunement, and being together with the AI, a concern that also appears, in different form, in research on social presence, rapport, and friendship-like bonding~\cite{merrill_jr_ai_2022, wienrich_long-term_2023}.

Related ideas appear across the literature under labels such as intimacy, friendship, psychological closeness, and extended self, which consistently signal stronger and more personally consequential human--technology relationships~\cite{banks_conceptualization_2026, carolus_smartphones_2019, ki_can_2020}. Considered together, these five indicators capture theoretically central and empirically differentiating features of MC as it unfolds as a motivated, reconstitutive, and effectual process.
\subsection{MC Wellbeing Outcomes: Need Satisfaction and Shifted Loneliness}

Human--AI relationships correspond with social wellbeing outcomes in positive, negative, and conflicting ways. Among the most examined of these is loneliness. On the beneficial side, AI companions can alleviate state loneliness by providing responsive conversational availability that leaves users feeling heard, making these systems highly useful for isolated individuals~\cite{de_freitas_ai_2026, merrill_jr_ai_2022}. However, these benefits are limited and can reverse depending on the user. Heavy, emotionally dependent engagement with social chatbots can displace human contact and exacerbate loneliness, particularly for users who already lack strong human social support~\cite{fang_how_2025, laestadius_too_2024, zhang_rise_2025}. Because the psychological effects of AI companions vary substantially based on age, attachment style, and usage patterns~\cite{ciriello_not_2026, liu_chatbot_2024, liu_heterogeneous_2025}, examining loneliness across distinct relational profiles may help clarify these divergent outcomes. This leads to our second research question:

\noindent\textbf{RQ2:} (How) are these human--AI companionship profiles associated with loneliness?

Alongside loneliness, human--AI relationships can directly impact RNS, the fundamental human experience of feeling connected, cared for, and socially significant~\cite{ryan_self-determination_2000}. Rather than remaining limited to instrumental tasks, technologies can directly fulfill basic psychological needs when they provide forms of social connection that users experience as authentically supportive and affirming~\cite{janssen_psychological_2024, machia_social_2024, moradbakhti_development_2024}. In practice, users frequently experience AI companions as accepting, nonjudgmental, and emotionally available, fostering deep trust and friendship-like bonds with clear social value~\cite{ki_can_2020, pentina_exploring_2023, skjuve_my_2021}. Importantly, however, artificial intelligence does not universally satisfy social needs. Relatedness experiences are highly variable and depend fundamentally on how a user interprets and structures the relationship, underscoring the need to examine these outcomes through multidimensional relationship profiles. This leads to our third research question:

\noindent\textbf{RQ3:} (How) are these human--AI companionship profiles associated with relatedness need satisfaction?

\section{Method}

\subsection{Data and Sample}


This study comprises a secondary analysis of data from a prior study \cite{banks_measuring_2026} in which the author developed and validated a self-report measure of machine companionship in interactions with AI companions. The data are publicly available at the project's online repository~\cite{banks_measuring_2025} and are used with permission. We limit analysis to the original scale-development sample (and not the validation data) because it is the larger of the two samples ($N = 467$ vs.\ 249) and therefore affords greater statistical power for the estimation of latent profiles. The original data were collected through an online survey from adults aged 18 years or older from the United States or the United Kingdom who were fluent or first-language English speakers and self-identified as interacting with text-based AI for social purposes; sample AI conversations were vetted by the original author to verify the latter requirement. The final analytic sample consisted of $N = 467$ participants and included users of general-purpose large language models, dedicated AI companion applications, and social media-embedded conversational AI~\cite{banks_measuring_2026}.

Demographics were reported in the source study~\cite{banks_measuring_2026} as follows. Participants were approximately evenly divided across binary sex categories and ranged in age from 19 to 76 years ($M = 38.30$, $SD = 12.04$). Education levels varied: 26.8\% held a high school diploma or less, 11.1\% an associate degree or some college, 39.8\% a bachelor's degree or equivalent, and 22.3\% a graduate degree. Most participants' named AI companions (AICs) were general-purpose LLMs (74.9\%; e.g., ChatGPT, Gemini), followed by dedicated companion apps (19.7\%; e.g., Replika, Kindroid) and social platform-embedded AI (5.4\%; e.g., Grok, Meta AI). Participants reported an average of 1.67 companions ($SD = 1.97$), approximately 5.5 weekly hours of interaction ($SD = 7.53$), and a mean duration since creation of the reported AI companion of 18.94 months ($SD = 31.08$).

\subsection{Measures}

Measures and their internal consistency were reported in the source study and are summarized here. All measures are 7-point Likert-style scales, unless otherwise indicated.

\textbf{Eudaimonic Exchange (EE) and Connective Coordination (CC).} The two primary dimensions of MC were measured using the 12-item MC Scale that emphasized one specific AI companion~\cite{banks_measuring_2026}. The scale includes two dimensions: the EE subscale comprises seven items assessing mutually elevating and purposeful interactions ($\omega = .93$), and the CC subscale comprises five items capturing interpersonal synchronization and bonding ($\omega = .92$).

\textbf{Mind Perception.} The extent to which users attributed experiential capacities to their AI companion was assessed using the 11-item experience dimension of the Mind Perception Scale~\cite{gray_dimensions_2007}. Participants rated their agreement that their AI companion is capable of a range of experiential states (e.g., pleasure, fear, consciousness, joy). The present analysis uses only the experience subscale, because the capacity to feel and be affected is the facet most directly implicated in relational closeness ($\omega = .96$).

\textbf{Inclusion of Other in the Self (IOS).} Perceived self--other overlap between the user and the AI was measured using an adapted version of the single-item Inclusion of Other in the Self measure~\cite{aron_inclusion_1992}. Participants were prompted to consider how they felt about their AI companion in the current moment and selected one of seven visual Venn diagrams, choosing the one that best represented the degree of overlap between themselves and their AI companion.

\textbf{Utilitarian Motivation.} The degree to which participants were driven to interact with their AI for practical and functional purposes was measured using an adapted 12-item utilitarian dimension from the consumer attitude scale of \citet{voss_measuring_2003}. Participants were asked, ``How much do each of the following motivate you to interact with your companion?'' and rated attributes such as ``effective,'' ``functional,'' ``useful,'' and ``productive'' for the degree to which they motivate the MC relationship. The scale demonstrated high internal consistency ($\omega = .95$).

\textbf{Loneliness.} Self-reported loneliness over the preceding two weeks was measured using an adaptation of the three-item UCLA Loneliness Scale~\cite{hughes_short_2004}. Participants indicated the frequency of specific feelings (lacking companionship, feeling left out, feeling isolated). The measure exhibited high internal consistency ($\omega = .94$).

\textbf{Relatedness Need Satisfaction.} Fulfillment of the fundamental human need for social connection was measured using an adapted version of the relatedness satisfaction dimension from the Basic Psychological Need Satisfaction and Frustration Scale~\cite{chen_basic_2015}. Participants evaluated items capturing the extent to which interacting with the AI influenced feelings of closeness and care (e.g., ``how close and connected you feel with others''). Responses demonstrated high internal consistency ($\omega = .94$).

\textbf{AI Companion Role.} Participants indicated the social role their named AI companion played in their life, using a multiple-choice item listing possible roles (acquaintance, friend, romantic partner, spouse, sibling, therapist, and mentor) with an open-field option for other descriptions~\cite{banks_measuring_2026}. Responses were classified by the researcher into four role categories: intimate (friend, romantic partner, spouse, or sibling), therapist/mentor (adviser, mentor, teacher, or therapist), assistant/collaborator (assistant, collaborator, or employee), and acquaintance. Non-relational responses were treated as missing. Thus, 433 participants (92.7\%) assigned to a relational role category (intimate, $n = 193$, 41.3\%; therapist/mentor, $n = 144$, 30.8\%; assistant/collaborator, $n = 40$, 8.6\%; acquaintance, $n = 56$, 12.0\%), and 34 (7.3\%) were missing.

\textbf{Average Weekly Use.} Participants reported their average weekly chat time with their AI companion. Responses were recorded in hours ($M = 5.50$, $SD = 7.53$) and used in this metric as a covariate in the outcome analyses.

\subsection{Analytical Strategy}

To identify distinct profiles of human--AI relationships (RQ1), we conducted latent profile analysis in Latent GOLD 6.0~\cite{vermunt_lg-syntax_2021}, estimating solutions of one to nine profiles from the five continuous indicators with 1,000 random start values to avoid local maxima. This upper bound was chosen conservatively, since prior person-centered studies of human--AI and human--technology relationships have identified relatively few subgroups, typically between two and seven~\cite{alabed_more_2024, liu_heterogeneous_2025, muller_chatbot_2019, schweitzer_servant_2019, smrke_aging_2022, toivonen_varieties_2024, wienrich_long-term_2023, yi_potential_2024}, so a nine-profile ceiling comfortably exceeds this range without constraining the optimal solution. The number of profiles was selected by comparing solutions on the Akaike and Bayesian information criteria (AIC, BIC; lower values indicating better fit), the Vuong--Lo--Mendell--Rubin likelihood-ratio test (VLMR; a significant value favoring the $k$- over the $(k-1)$-profile solution), and entropy (classification precision), alongside the interpretability and distinctiveness of the profiles~\cite{nylund_deciding_2007, spurk_latent_2020}.

To characterize the retained profiles, we examined whether participants' assigned AI companion role predicted profile membership, using the bias-corrected three-step (Step-3) approach in Latent GOLD with maximum-likelihood correction~\cite{bakk_robustness_2016, vermunt_latent_2010}. Role was entered as three dummy variables (Intimate, Therapist/Mentor, and Assistant/Collaborator) with Acquaintance as the reference category and Profile 1 as the reference profile. This model was estimated on the 433 participants who were assigned a relational role.

Using the same three-step approach, we then examined how the profiles differed in loneliness and RNS (RQ2, RQ3). Because latent profile analysis was estimated on the full sample but the outcome models required complete data on both the outcome and the covariate, the loneliness analysis was conducted on 447 participants and the RNS analysis on 448 (19 participants were missing weekly-use values, and 1 was additionally missing a loneliness value). Average weekly use was included as a covariate in both outcome models, because usage intensity varies across relational profiles~\cite{liu_chatbot_2024, wienrich_long-term_2023} and is associated with users' social and emotional outcomes~\cite{fang_how_2025}. This allowed differences across profiles in loneliness and RNS to be estimated independently of the amount of time spent interacting with the AI.

\section{Results}

\subsection{Latent Human--AI Relationship Profiles (RQ1)}

Latent profile models were estimated for one- to nine-profile solutions. The AIC was lowest at nine profiles, but the BIC was lowest at six, and the VLMR test was significant through six profiles but not at seven ($p = .17$), indicating that a seventh profile did not improve fit. Fit statistics for all solutions are reported in Table~\ref{tab:fit}. We compared the six-profile solution against a more parsimonious five-profile solution on interpretability, distinctiveness, and classification quality. The six-profile solution was not retained because its additional profile did not form a distinct relational pattern but split existing groups, adding no conceptual novelty and reducing the distinctiveness of the other profiles. The five-profile solution produced clearly differentiated, interpretable profiles with strong classification precision (entropy $= .86$, classification error $= .08$).  We therefore retained the five-profile solution.

\begin{table*}[t]
\caption{Model fit statistics for the one- to nine-profile solutions. The retained five-profile solution is shown in bold.}
\Description{A nine-row table of latent profile model fit statistics, one row per solution from one to nine profiles. Columns report log-likelihood, BIC, AIC, number of parameters, maximum bivariate residual, the Vuong-Lo-Mendell-Rubin likelihood ratio test statistic and its p-value, classification error, and entropy. AIC decreases monotonically across solutions and is lowest at nine profiles; BIC is lowest at six profiles; the VLMR test is significant through six profiles but not at seven. The five-profile row is bolded to indicate the retained solution, with entropy of .862 and classification error of .079.}
\label{tab:fit}
\small
\begin{tabular}{@{}l r r r r r r r r r@{}}
\toprule
\textbf{Model} & \textbf{LL} & \textbf{BIC(LL)} & \textbf{AIC(LL)} & \textbf{\emph{N}\textsubscript{par}} & \textbf{Max.\ BVR} & \textbf{VLMR} & \textbf{\emph{p}} & \textbf{Class.\ err.} & \textbf{Entropy \emph{R}\textsuperscript{2}} \\
\midrule
1-Profile & $-4131.35$ & 8324.16 & 8282.70 & 10 & 229.89 & --- & --- & .000 & 1.000 \\
2-Profile & $-3704.20$ & 7537.47 & 7450.40 & 21 & 81.60 & 854.29 & $<.001$ & .043 & .845 \\
3-Profile & $-3511.59$ & 7219.86 & 7087.18 & 32 & 50.21 & 385.21 & $<.001$ & .071 & .836 \\
4-Profile & $-3429.70$ & 7123.70 & 6945.41 & 43 & 28.99 & 163.77 & $<.001$ & .080 & .842 \\
\textbf{5-Profile} & $\bm{-3327.23}$ & \textbf{6986.37} & \textbf{6762.46} & \textbf{54} & \textbf{15.71} & \textbf{204.94} & $\bm{<.001}$ & \textbf{.079} & \textbf{.862} \\
6-Profile & $-3192.25$ & 6784.02 & 6514.51 & 65 & 13.19 & 269.95 & $<.001$ & .100 & .855 \\
7-Profile & $-3159.67$ & 6786.46 & 6471.34 & 76 & 10.40 & 65.16 & .177 & .112 & .848 \\
8-Profile & $-3125.69$ & 6786.11 & 6425.38 & 87 & 7.20 & 67.96 & .009 & .098 & .862 \\
9-Profile & $-3098.33$ & 6799.00 & 6392.66 & 98 & 5.40 & 54.71 & .060 & .116 & .852 \\
\bottomrule
\end{tabular}

\vspace{2pt}
\footnotesize
\emph{Note.} LL = log-likelihood; BIC(LL) = Bayesian information criterion based on log-likelihood; AIC(LL) = Akaike information criterion based on log-likelihood; \emph{N}\textsubscript{par} = number of parameters; Max.\ BVR = maximum bivariate residual; VLMR = Vuong--Lo--Mendell--Rubin likelihood ratio test; Class.\ err.\ = classification error; Entropy \emph{R}\textsuperscript{2} = entropy-based pseudo \emph{R}\textsuperscript{2}.
\end{table*}

\begin{figure*}[t]
\centering
\includegraphics[width=\linewidth]{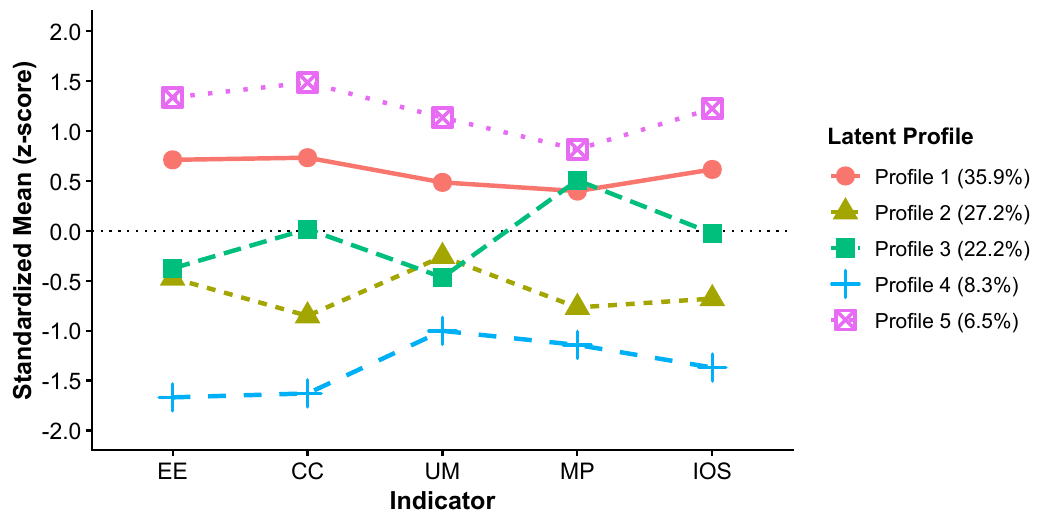}
\caption{Standardized indicator means for the five latent profiles. Mind Perception shows the narrowest spread across profiles, and even the highest-scoring profile remains below the scale midpoint in raw terms.}
\Description{A line plot with five indicators on the horizontal axis (EE, CC, UM, MP, IOS) and standardized z-scores from -2.0 to 2.0 on the vertical axis. Each profile is drawn as a distinct line with its own colour, line style, and point shape. Profile 5 (intimate-bonded) runs highest throughout, between roughly 0.8 and 1.5. Profile 1 (intimate-casual) is next, staying between roughly 0.4 and 0.7. Profile 4 (objectified-utilitarian) runs lowest throughout, between roughly -1.7 and -1.0. Profiles 2 and 3 cross each other in the middle of the range: Profile 3 (neutral-utilitarian) starts near -0.4, rises to the sample mean at CC, dips at UM, then returns to the mean at IOS, while Profile 2 (disengaged-utilitarian) dips to -0.8 at CC before levelling off near -0.7. The profiles converge most closely on Utilitarian Motivation and on Mind Perception, and spread most widely on Connective Coordination and Inclusion of Other in the Self.}
\label{fig:profiles}

\vspace{2pt}
\footnotesize
\emph{Note.} Standardized means (z-scores) for the five latent profiles across the five indicators ($N = 467$). Z-scores were computed using the overall sample mean and standard deviation for each indicator. The horizontal reference line at $z = 0$ represents the overall sample mean. Percentages in the legend indicate the proportion of the sample assigned to each profile. EE = Eudaimonic Exchange; CC = Connective Coordination; UM = Utilitarian Motivation; MP = Mind Perception (experience dimension); IOS = Inclusion of Other in the Self.
\end{figure*}

We interpret the profiles as differing qualitatively in the formality and intensity of the relational orientation. Profile~1 (intimate-casual; 35.9\%) scored above the sample mean on all five indicators, most markedly on Connective Coordination and Inclusion of Other in the Self. Profile~2 (disengaged-utilitarian; 27.2\%) scored below the mean on all five, most markedly on Connective Coordination, Mind Perception, and Inclusion of Other in the Self, with Utilitarian Motivation only slightly below average. Profile~3 (neutral-utilitarian; 22.2\%) showed a distinct configuration, near the mean on Connective Coordination and Inclusion of Other in the Self, slightly below on Eudaimonic Exchange and Utilitarian Motivation, and above it on Mind Perception, the second-highest value of any profile though still near the midpoint (3.43). Profile~4 (objectified-utilitarian; 8.3\%) scored more than one standard deviation below the mean on every indicator, with Inclusion of Other in the Self at the floor. Profile~5 (intimate-bonded; 6.5\%) scored highest of all profiles on every indicator, reaching the upper end of the scale on the relational and utilitarian measures while its Mind Perception remained near the midpoint (3.94). Table~\ref{tab:profiles} reports profile prevalences and indicator means (on the original 1--7 scale). When the profiles are mapped as standardized scores relative to the sample mean (Figure~\ref{fig:profiles}), an intuitive gradient emerges, one based on the degree to which one has high or low levels of the profile variables (generally: Profile 5 $>$ 1 $>$ 3 $>$ 2 $>$ 4).

\begin{table*}[t]
\caption{Profile prevalences and indicator means for the retained five-profile solution.}
\Description{A table with five indicator rows and one profile-size row, crossed against the five profiles and an overall column. Profile sizes range from 35.9 percent for Profile 1 to 6.5 percent for Profile 5. Across all five indicators, Profile 5 (intimate-bonded) has the highest means and Profile 4 (objectified-utilitarian) the lowest, with the remaining profiles falling between them. Utilitarian Motivation is the highest-scoring indicator in every profile, ranging from 4.56 to 6.95. Mind Perception remains at or below the scale midpoint of 4 in every profile, with a maximum of 3.94.}
\label{tab:profiles}
\small
\begin{tabular}{@{}l r r r r r r@{}}
\toprule
& \textbf{Profile 1} & \textbf{Profile 2} & \textbf{Profile 3} & \textbf{Profile 4} & \textbf{Profile 5} & \\
\textbf{Indicator} & \textbf{Intimate-} & \textbf{Disengaged-} & \textbf{Neutral-} & \textbf{Objectified-} & \textbf{Intimate-} & \textbf{Overall} \\
& \textbf{casual} & \textbf{utilitarian} & \textbf{utilitarian} & \textbf{utilitarian} & \textbf{bonded} & \\
\midrule
Profile size (\%) & 35.9 & 27.2 & 22.2 & 8.3 & 6.5 & 100.0 \\
\addlinespace
Eudaimonic Exchange & 6.19 & 4.76 & 4.90 & 3.28 & 6.95 & 5.32 \\
Connective Coordination & 5.46 & 2.87 & 4.28 & 1.57 & 6.69 & 4.25 \\
Utilitarian Motivation & 6.23 & 5.40 & 5.21 & 4.56 & 6.95 & 5.69 \\
Mind Perception & 3.36 & 1.69 & 3.43 & 1.13 & 3.94 & 2.78 \\
Inclusion of Other in the Self & 4.58 & 2.28 & 3.38 & 1.00 & 5.69 & 3.46 \\
\bottomrule
\end{tabular}

\vspace{2pt}
\footnotesize
\emph{Note.} $N = 467$. Indicator values are estimated means on the original 1--7 response scale. Profile size is the percentage of the sample assigned to each profile; for that row, Overall is the total, and for the indicator rows, the Overall column reports the sample-wide mean of each indicator. Profile labels summarize the configuration of indicator means. Mind Perception remains at or below the scale midpoint (4.0) in every profile, including the most companionate one.
\end{table*}

If we then map those profiles to the role assigned to the AI companion, we observe that the AI companion role predicts profile membership. With acquaintance as the reference role and Profile~1 as the reference profile, intimate roles ($\mathit{Wald} = 23.12$, $p < .001$) and assistant/collaborator roles ($\mathit{Wald} = 11.61$, $p = .020$) were significantly associated with membership, whereas therapist/mentor roles were not ($\mathit{Wald} = 9.10$, $p = .059$). Intimate roles became progressively more common across the gradient, from Profile~4 (28\%) and Profile~2 (29\%), through Profile~3 (35\%), to Profiles~1 (61\%) and 5 (65\%). Functional (assistant/collaborator) roles showed the reverse pattern, most common in Profile~4 (28\%) and Profile~2 (16\%) and nearly absent from Profile~5 ($<1$\%). Acquaintance roles were concentrated in the two middle profiles (Profile~2, 19\%; Profile~3, 23\%), and therapist/mentor roles were relatively evenly distributed across profiles (30--37\%).

Altogether, we interpret these findings to indicate there are discrete machine companionship profiles, and those profiles appear to represent an underlying continuum. The cohering variables alongside their associated roles suggest this continuum is a gradient of self-expansion~\cite{aron_self-expansion_2022}, the extent to which one engages the AI companion to satisfy a fundamental drive to increase their own material and social resources, knowledge and perspectives, and possible identities.

\subsection{Association of Latent Profiles with Loneliness (RQ2)}

Profile membership was significantly associated with loneliness, $\mathit{Wald}(4) = 13.99$, $p = .007$, though the profiles accounted for only a small proportion of variance ($R^2 = .04$). Average weekly use, included as a covariate, was not significantly associated with loneliness ($p = .063$). Estimated mean loneliness differed across the five profiles (overall $M = 3.52$): Profile~4 reported the highest loneliness ($M = 4.05$), followed by Profile~3 ($M = 3.92$), Profile~2 ($M = 3.40$), and Profile~1 ($M = 3.39$), and Profile~5 reported the lowest ($M = 2.63$). With Profile~5 as the reference, Profile~3 ($b = 1.29$, $\mathit{SE} = 0.41$, $z = 3.13$, $p = .002$) and Profile~4 ($b = 1.42$, $\mathit{SE} = 0.49$, $z = 2.90$, $p = .004$) reported significantly higher loneliness, whereas Profile~2 ($p = .062$) and Profile~1 ($p = .061$) did not differ significantly from it. Full estimates are reported in Table~\ref{tab:loneliness}. Pairwise comparisons across all profile pairs (Table~\ref{tab:pairwise}) further clarified these differences: Profile~5 reported significantly lower loneliness than both Profile~3 ($p = .002$) and Profile~4 ($p = .004$), and Profile~1 differed significantly from Profile~3 ($p = .030$). No other comparison reached significance (all $p > .05$), indicating that most profiles did not differ in reported loneliness.

\begin{table}[t]
\caption{Regression of loneliness on profile membership, with Profile~5 (intimate-bonded) as the reference category.}
\Description{A regression table with seven predictor rows. The intercept is 2.50. The overall profile effect is significant with a Wald statistic of 13.99 on 4 degrees of freedom and p equal to .007. Relative to Profile 5, Profile 4 has a coefficient of 1.42 and Profile 3 a coefficient of 1.29, both significant. Profiles 2 and 1 have coefficients of 0.77 and 0.76 respectively, both marginally non-significant at p just above .06. Average weekly use, included as a covariate, has a coefficient of 0.02 and is not significant.}
\label{tab:loneliness}
\small
\begin{tabular}{@{}l r r r r r r@{}}
\toprule
\textbf{Predictor} & \textbf{\emph{b}} & \textbf{\emph{SE}} & \textbf{\emph{z}} & \textbf{\emph{p}} & \textbf{Wald} & \textbf{\emph{df}} \\
\midrule
Intercept & 2.50 & 0.38 & 6.62 & $<.001$ & 43.89 & 1 \\
\addlinespace
Profile (overall) & & & & .007 & 13.99 & 4 \\
\quad Profile 4 & 1.42 & 0.49 & 2.90 & .004 & --- & --- \\
\quad Profile 2 & 0.77 & 0.41 & 1.86 & .062 & --- & --- \\
\quad Profile 3 & 1.29 & 0.41 & 3.13 & .002 & --- & --- \\
\quad Profile 1 & 0.76 & 0.40 & 1.87 & .061 & --- & --- \\
\quad Profile 5 (reference) & 0 & --- & --- & --- & --- & --- \\
\addlinespace
Average weekly use & 0.02 & 0.01 & 1.86 & .063 & 3.47 & 1 \\
\bottomrule
\end{tabular}

\vspace{2pt}
\footnotesize
\emph{Note.} $N = 447$. Outcome variable = loneliness. \emph{b} = unstandardized regression coefficient; \emph{SE} = standard error; \emph{z} = \emph{z}-statistic; Wald = Wald chi-square statistic; \emph{df} = degrees of freedom. Profile~5 was set as the reference category. The Wald statistic for Profile (overall) tests the joint significance of all profile contrasts. Dashes indicate values not estimated for the reference category.
\end{table}
\subsection{Association of Latent Profiles with Relatedness Need Satisfaction (RQ3)}

Relatedness need satisfaction was regressed on profile membership, again including average weekly use as a covariate. Profile membership was strongly associated with RNS, $\mathit{Wald}(4) = 135.45$, $p < .001$, accounting for a substantial proportion of variance ($R^2 = .34$); average weekly use was not significantly associated with the outcome ($p = .380$). Estimated mean RNS differed across the five profiles (overall $M = 4.20$): Profile~5 reported the highest ($M = 6.00$), followed by Profile~1 ($M = 5.10$), Profile~3 ($M = 4.29$), Profile~2 ($M = 3.05$), and Profile~4 ($M = 2.67$). With Profile~5 as the reference, all four remaining contrasts were significant (all $p < .001$): Profiles~1, 2, 3, and 4 all reported lower RNS than Profile~5, with the difference increasing down the gradient, from Profile~1 ($b = -0.90$) through Profile~3 ($b = -1.72$) and Profile~2 ($b = -2.96$) to Profile~4 ($b = -3.34$). Full estimates are reported in Table~\ref{tab:rns}. Pairwise comparisons (Table~\ref{tab:pairwise}) supported these differences: nine of the ten comparisons were significant (all $p < .001$), the only exception being Profile~2 versus Profile~4 ($p = .290$). In sum, the five profiles differed significantly in RNS, with Profile~5 reporting the highest level and Profile~4 the lowest, and a clear ordering among the profiles; only Profiles~2 and 4 did not differ significantly.

\begin{table}[t]
\caption{Pairwise comparisons of loneliness and relatedness need satisfaction across the five latent profiles.}
\Description{A table of ten pairwise profile comparisons, with Wald statistics, degrees of freedom, and p-values reported separately for loneliness and for relatedness need satisfaction. For loneliness, only three of the ten comparisons reach significance: Profile 1 versus Profile 3, Profile 3 versus Profile 5, and Profile 4 versus Profile 5. For relatedness need satisfaction, nine of the ten comparisons are significant at p below .001, the sole exception being Profile 2 versus Profile 4 at p equal to .290. All comparisons have one degree of freedom.}
\label{tab:pairwise}
\small
\begin{tabular}{@{}l r r r r r r@{}}
\toprule
& \multicolumn{3}{c}{\textbf{Loneliness}} & \multicolumn{3}{c}{\textbf{Relatedness Need Sat.}} \\
\cmidrule(lr){2-4} \cmidrule(l){5-7}
\textbf{Comparison} & \textbf{Wald} & \textbf{\emph{df}} & \textbf{\emph{p}} & \textbf{Wald} & \textbf{\emph{df}} & \textbf{\emph{p}} \\
\midrule
Profile 1 vs.\ 2 & 0.01 & 1 & .960 & 67.09 & 1 & $<.001$ \\
Profile 1 vs.\ 3 & 4.70 & 1 & .030 & 19.97 & 1 & $<.001$ \\
Profile 1 vs.\ 4 & 3.53 & 1 & .060 & 53.72 & 1 & $<.001$ \\
Profile 1 vs.\ 5 & 3.51 & 1 & .061 & 11.27 & 1 & $<.001$ \\
Profile 2 vs.\ 3 & 3.47 & 1 & .062 & 20.55 & 1 & $<.001$ \\
Profile 2 vs.\ 4 & 3.09 & 1 & .079 & 1.09 & 1 & .290 \\
Profile 2 vs.\ 5 & 3.48 & 1 & .062 & 78.64 & 1 & $<.001$ \\
Profile 3 vs.\ 4 & 0.14 & 1 & .710 & 22.85 & 1 & $<.001$ \\
Profile 3 vs.\ 5 & 9.81 & 1 & .002 & 37.22 & 1 & $<.001$ \\
Profile 4 vs.\ 5 & 8.43 & 1 & .004 & 70.15 & 1 & $<.001$ \\
\bottomrule
\end{tabular}

\vspace{2pt}
\footnotesize
\emph{Note.} Pairwise comparisons of estimated mean levels of loneliness and relatedness need satisfaction (RNS) across the five latent profiles, controlling for average weekly AI use. Wald = Wald chi-square statistic; \emph{df} = degrees of freedom.
\end{table}
\begin{table}[t]
\caption{Regression of relatedness need satisfaction on profile membership, with Profile~5 (intimate-bonded) as the reference category.}
\Description{A regression table with seven predictor rows. The intercept is 5.96. The overall profile effect is highly significant, with a Wald statistic of 135.44 on 4 degrees of freedom. Relative to Profile 5, all four remaining profiles show significantly lower relatedness need satisfaction, and the size of the deficit increases down the continuum: Profile 1 is 0.90 lower, Profile 3 is 1.72 lower, Profile 2 is 2.96 lower, and Profile 4 is 3.34 lower, all at p below .001. Average weekly use, included as a covariate, has a coefficient of 0.01 and is not significant.}
\label{tab:rns}
\small
\begin{tabular}{@{}l r r r r r r@{}}
\toprule
\textbf{Predictor} & \textbf{\emph{b}} & \textbf{\emph{SE}} & \textbf{\emph{z}} & \textbf{\emph{p}} & \textbf{Wald} & \textbf{\emph{df}} \\
\midrule
Intercept & 5.96 & 0.27 & 22.48 & $<.001$ & 505.21 & 1 \\
\addlinespace
Profile (overall) & & & & $<.001$ & 135.44 & 4 \\
\quad Profile 4 & $-3.34$ & 0.40 & $-8.38$ & $<.001$ & --- & --- \\
\quad Profile 2 & $-2.96$ & 0.33 & $-8.87$ & $<.001$ & --- & --- \\
\quad Profile 3 & $-1.72$ & 0.28 & $-6.10$ & $<.001$ & --- & --- \\
\quad Profile 1 & $-0.90$ & 0.27 & $-3.36$ & $<.001$ & --- & --- \\
\quad Profile 5 (reference) & 0 & --- & --- & --- & --- & --- \\
\addlinespace
Average weekly use & 0.01 & 0.01 & 0.88 & .380 & 0.77 & 1 \\
\bottomrule
\end{tabular}

\vspace{2pt}
\footnotesize
\emph{Note.} $N = 448$. Outcome variable = relatedness need satisfaction. \emph{b} = unstandardized regression coefficient; \emph{SE} = standard error; \emph{z} = \emph{z}-statistic; Wald = Wald chi-square statistic; \emph{df} = degrees of freedom. Profile~5 was set as the reference category. The Wald statistic for Profile (overall) tests the joint significance of all profile contrasts. Dashes indicate values not estimated for the reference category.
\end{table}

\section{Discussion}

\begin{figure*}[t]
\centering
\includegraphics[width=\linewidth]{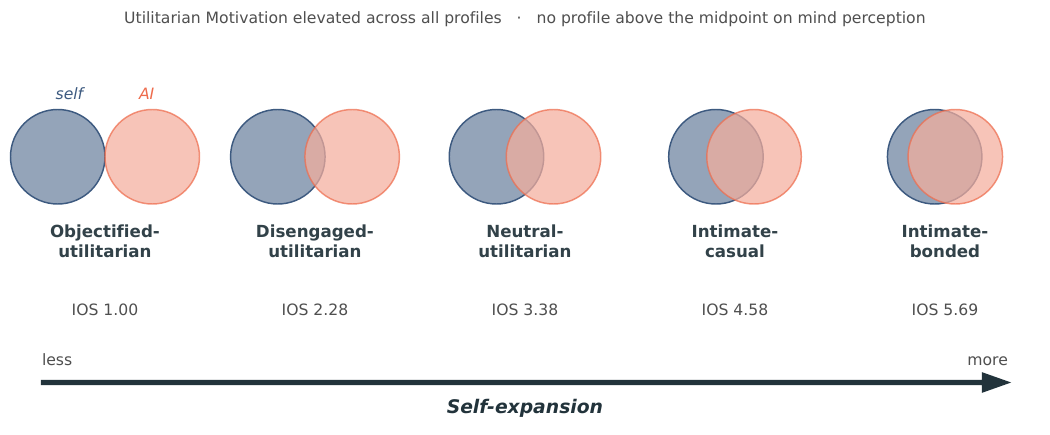}
\caption{The five profiles arranged along the underlying gradient of self-expansion, from least to most self-expanded. Circle overlap represents each profile's mean IOS score.}
\Description{Five pairs of overlapping circles arranged left to right along an arrow labelled self-expansion, running from less to more. In each pair a blue circle represents the self and an orange circle the AI. The objectified-utilitarian profile at the left shows two circles that touch without overlapping; the overlap increases through the disengaged-utilitarian, neutral-utilitarian, and intimate-casual profiles, until the intimate-bonded profile at the right shows the circles largely superimposed. Each pair is labelled with its profile name and mean IOS score. A line above the diagram notes that utilitarian motivation is elevated across all profiles and that no profile scores above the midpoint on mind perception.}
\label{fig:selfexpansion}

\vspace{2pt}
\footnotesize
\emph{Note.} Profiles are arranged from left to right by increasing self-expansion as indicated by the arrow. Circle overlap is drawn in proportion to these IOS means. The blue circle represents the self and the orange circle the AI companion. The annotation above the diagram summarizes two patterns across all profiles.
\end{figure*}

The current study identified five latent profiles of human--AI relationships based on five indicators of relational experience (RQ1) and examined their associations with loneliness (RQ2) and relatedness need satisfaction (RQ3). The five profiles emerged as follows: objectified-utilitarian (8.3\%), disengaged-utilitarian (27.2\%), neutral-utilitarian (22.2\%), intimate-casual (35.9\%), and intimate-bonded (6.5\%). The distribution of indicators across these profiles suggests they differ qualitatively in how relational facets converge. Furthermore, their correspondence with assigned AI role labels indicates that these profiles occupy distinct positions along an underlying gradient of self-expansion (Figure~\ref{fig:selfexpansion}).


Self-expansion theory posits that individuals possess a fundamental desire to expand the self, defined as the drive to ``increase their self-efficacy, perspectives, competence, and resources''~\cite[p.~3821]{aron_self-expansion_2022}. This theoretical framework cohesively unifies the variables in our profiles. IOS serves as the literal operationalization of this expansion, and it tracked cleanly upward across our gradient. Furthermore, the model distinguishes between expansion pursued for instrumental gain and expansion valued for the affective experience itself~\cite{aron_inclusion_1992}. Our data show these motives are not mutually exclusive; instead, instrumental and companionate orientations did not trade off but rose and fell together. Utilitarian motivation was elevated across all profiles, indicating a baseline drive to acquire resources and information; this is likely reflective of the high proportion of participants who socially engage general-purpose LLMs (rather than dedicated AI companion apps). Moving up the gradient, this functional utility increasingly co-occurred with EE and CC, which represent the affective, companionate rewards of the expansion process. In other words, deep human--AI bonds are strong versions of a single orientation where tool use and social connection operate together to expand the user's self-concept. 

The self-expansion framework also clarifies our findings regarding mind perception. Because self-expansion is most often studied within human relationships, interpersonal closeness is typically assumed to involve another mind. In this study, however, no profile exceeded the scale midpoint on mind perception. Even the intimate-bonded profile averaged an ambivalent 3.94, and the least-engaged profiles scored near the floor. This indicates that self-expansion does not require users to perceive the AI as a sentient agent capable of feeling. This finding aligns with the broader scope of the self-expansion model, which demonstrates that individuals can include non-human targets, such as nature, God, technologies, and brands, in their self-concept~\cite{aron_self-expansion_2022}. Consequently, consistent with \citet{norskov_sociomorphing_2020}, users appear to sociomorph the machine, utilizing its interactive and responsive capacities to facilitate their own expansion without projecting human consciousness onto it.

These profiles also correspond distinctly with wellbeing outcomes. Profile membership was strongly associated with RNS which rose steadily along the continuum. Almost all profiles differed from one another in RNS, the exception being the two least expanded profiles, whose CC fell below the midpoint. CC therefore appears to mark where these relationships begin to satisfy relatedness needs, with users in the more expanded profiles reporting a greater sense of care and closeness through the AI relationship. Loneliness, by contrast, was only weakly associated with profile membership. Because RNS measures the relationship-specific experience while loneliness captures broader social isolation, this divergence suggests the benefits of AI companionship are specific to the human-AI social domain. AI can fulfill immediate relational needs but may not inherently resolve systemic human disconnection, aligning with findings that the effects of AI on loneliness vary heavily across users~\cite{fang_how_2025, liu_heterogeneous_2025}.

\subsection{Implications for Theory and Design}

\textbf{Theoretical Implications.} The present findings offer several theoretical implications. First, whereas prior work typically categorizes AI users based on isolated variables like assigned roles or traits~\cite{schweitzer_servant_2019, wienrich_long-term_2023}, our profiling reveals that people relate to AI along a continuous gradient of self-expansion~\cite{aron_self-expansion_2022}. Second, instrumental and companionate orientations co-occur. This supports recent evidence that human--AI interactions are highly fluid, blurring the boundaries between productivity assistants and emotional companions~\cite{manoli_2026_digital}. Consequently, this co-occurrence departs from traditional frameworks that strictly separate transactional agents from social chatbots~\cite{clark_2019_what}. Third, users achieved deep companionship without firmly attributing a conscious mind to the AI. This aligns with bounded personhood, wherein users form attachments while maintaining ontological skepticism~\cite{manoli_2026_digital}. It also corroborates findings that users evaluate AI based on pragmatic interactional cues rather than abstract sentience~\cite{schimmelpfennig_2026_humanlike}, supporting sociomorphing frameworks where engagement relies on interactional coordination rather than projected human consciousness~\cite{norskov_sociomorphing_2020}. Finally, the strong association with relatedness need satisfaction alongside a weak association with general loneliness suggests domain-specific benefits, confirming that AI can fulfill immediate relational needs but does not inherently resolve systemic human disconnection~\cite{fang_how_2025, liu_heterogeneous_2025, manoli_2026_digital}.

\textbf{Implications for Design.} These theoretical insights translate into practical design guidelines. Because functional utility and social intimacy operate synergistically, developers might avoid strictly bifurcating systems into task versus chat modes. Interfaces can instead accommodate fluid use by blending practical affordances with relational cues~\cite{manoli_2026_digital}. Furthermore, since deep companionship occurs without strong mind attribution, designers need not over-engineer deceptive human-like emotions. Resources are better spent enhancing interactional competence, such as conversational flow and reliable responsiveness, which drive connection more effectively than claims of human-likeness~\cite{schimmelpfennig_2026_humanlike}. Finally, recognizing that AI fulfills specific relational needs rather than curing global loneliness helps ethically calibrate system framing. Platforms might position conversational AI as targeted relational supplements, utilizing reflective boundaries to support healthy self-expansion while mitigating overreliance~\cite{manoli_2026_digital}.


\subsection{Limitations and Future Directions}

Several limitations qualify these conclusions and point toward future directions. First, the current study leveraged data from a cross-sectional study; therefore, we cannot establish the direction of the observed associations and were limited by data collected in the source study. Future longitudinal designs are needed to clarify how these human--AI relationship profiles and their correlates unfold over time, and perhaps how members of profiles realize differential wellbeing and identity outcomes related to self-expansion. Second, that sourced sample consisted of self-selected, active users of text-based AI in the US and UK, which bounds generalizability. Future studies should evaluate whether these latent profiles generalize to more diverse populations and in voice-based or embodied contexts. Third, the engaged measurement of mind perception utilized only the experiential mental-capacities dimension; future work should continue exploring how varying perceptions of machines' agentic mental capacities impact self-expansion. Finally, loneliness was measured only in general terms, without distinguishing between disconnection from people and disconnection from the AI, an ambiguity that may help explain its weak association with the profiles and that future research should address.

\section{Conclusion}

This study finds that human--AI relationships are best understood as situated along a continuum of self-expansion. By profiling motivational, cognitive, and experiential indicators, we found that functional utility and social connectedness operate together as dimensions along a gradient signaling different strategies or preferences for expanding one's own social or utilitarian resources. Users engage conversational agents to expand their own capabilities and social experiences, doing so without needing to perceive the AI as a conscious, minded entity. Furthermore, as users move upward along this continuum, their relatedness needs are increasingly satisfied, though general loneliness remains largely unaffected. Ultimately, conceptualizing machine companionship as a continuous process of self-expansion provides a nuanced framework for understanding the mechanics, design possibilities, and specific psychological boundaries of human--AI connections.

\bibliographystyle{ACM-Reference-Format}
\bibliography{references}

@article{rogge_defining_2023,
	title = {Defining, designing and distinguishing artificial companions: {A} systematic literature review},
	volume = {15},
	number = {9},
	journal = {International Journal of Social Robotics},
	publisher = {Springer},
	author = {Rogge, Ayanda},
	year = {2023},
	pages = {1557--1579},
}

@article{strohmann_toward_2023,
	title = {Toward a design theory for virtual companionship},
	volume = {38},
	number = {3-4},
	journal = {Human–Computer Interaction},
	publisher = {Taylor \& Francis},
	author = {Strohmann, Timo and Siemon, Dominik and Khosrawi-Rad, Bijan and Robra-Bissantz, Susanne},
	year = {2023},
	pages = {194--234},
}

@article{skjuve_my_2021,
	title = {My chatbot companion-a study of human-chatbot relationships},
	volume = {149},
	journal = {International Journal of Human-Computer Studies},
	publisher = {Elsevier},
	author = {Skjuve, Marita and Følstad, Asbjørn and Fostervold, Knut Inge and Brandtzaeg, Petter Bae},
	year = {2021},
	pages = {102601},
}

@article{schweitzer_servant_2019,
	title = {Servant, friend or master? {The} relationships users build with voice-controlled smart devices},
	volume = {35},
	number = {7-8},
	journal = {Journal of Marketing Management},
	publisher = {Taylor \& Francis},
	author = {Schweitzer, Fiona and Belk, Russell and Jordan, Werner and Ortner, Melanie},
	year = {2019},
	pages = {693--715},
}

@article{chaturvedi_social_2023,
	title = {Social companionship with artificial intelligence: {Recent} trends and future avenues},
	volume = {193},
	journal = {Technological Forecasting and Social Change},
	publisher = {Elsevier},
	author = {Chaturvedi, Rijul and Verma, Sanjeev and Das, Ronnie and Dwivedi, Yogesh K.},
	year = {2023},
	pages = {122634},
}

@article{pentina_consumermachine_2023,
	title = {Consumer–machine relationships in the age of artificial intelligence: {Systematic} literature review and research directions},
	volume = {40},
	number = {8},
	journal = {Psychology \& Marketing},
	publisher = {Wiley Online Library},
	author = {Pentina, Iryna and Xie, Tianling and Hancock, Tyler and Bailey, Ainsworth},
	year = {2023},
	pages = {1593--1614},
}

@article{banks_measuring_2026,
	title = {Measuring {Machine} {Companionship} {Experiences}: {Scale} {Development} and {Validation} for {AI} {Companions}},
	journal = {Computers in Human Behavior},
	publisher = {Elsevier},
	author = {Banks, Jaime},
	year = {2026},
	pages = {108945},
}

@article{banks_conceptualization_2026,
	title = {Conceptualization, operationalization, and measurement of machine companionship: a scoping review},
	volume = {31},
	number = {2},
	journal = {Journal of Computer-Mediated Communication},
	publisher = {Oxford University Press},
	author = {Banks, Jaime and Li, Zhixin},
	year = {2026},
	pages = {zmaf027},
}

@article{alabed_more_2024,
	title = {More than just a chat: {A} taxonomy of consumers’ relationships with conversational {AI} agents and their well-being implications},
	volume = {58},
	number = {2},
	journal = {European Journal of Marketing},
	publisher = {Emerald Publishing Limited},
	author = {Alabed, Amani and Javornik, Ana and Gregory-Smith, Diana and Casey, Rebecca},
	year = {2024},
	pages = {373--409},
}

@article{tschopp_servant_2023,
	title = {Servant by default? {How} humans perceive their relationship with conversational {AI}},
	volume = {17},
	number = {3},
	journal = {Cyberpsychology: Journal of Psychosocial Research on Cyberspace},
	author = {Tschopp, Marisa and Gieselmann, Miriam and Sassenberg, Kai},
	year = {2023},
}

@article{ryan_self-determination_2000,
	title = {Self-determination theory and the facilitation of intrinsic motivation, social development, and well-being.},
	volume = {55},
	number = {1},
	journal = {American psychologist},
	publisher = {American Psychological Association},
	author = {Ryan, Richard M. and Deci, Edward L.},
	year = {2000},
	pages = {68},
}

@inproceedings{liu_heterogeneous_2025,
	title = {The heterogeneous effects of {AI} companionship: {An} empirical model of chatbot usage and loneliness and a typology of user archetypes},
	volume = {8},
	isbn = {3065-8365},
	number = {2},
	booktitle = {Proceedings of the {AAAI}/{ACM} {Conference} on {AI}, {Ethics}, and {Society}},
	author = {Liu, Auren R. and Pataranutaporn, Pat and Maes, Pattie},
	year = {2025},
	pages = {1585--1597},
}

@article{turkle_relational_2006,
	title = {Relational artifacts with children and elders: the complexities of cybercompanionship},
	volume = {18},
	number = {4},
	journal = {Connection Science},
	publisher = {Taylor \& Francis},
	author = {Turkle, Sherry and Taggart, Will and Kidd, Cory D. and Dasté, Olivia},
	year = {2006},
	pages = {347--361},
}

@article{nass_machines_2000,
	title = {Machines and mindlessness: {Social} responses to computers},
	volume = {56},
	number = {1},
	journal = {Journal of social issues},
	publisher = {Wiley Online Library},
	author = {Nass, Clifford and Moon, Youngme},
	year = {2000},
	pages = {81--103},
}

@article{gambino_building_2020,
	title = {Building a stronger {CASA}: {Extending} the computers are social actors paradigm},
	volume = {1},
	journal = {Human-Machine Communication},
	publisher = {Communication and Social Robotics Labs Kalamazoo, Michigan},
	author = {Gambino, Andrew and Fox, Jesse and Ratan, Rabindra A.},
	year = {2020},
	pages = {71--85},
}

@article{epley_seeing_2007,
	title = {On seeing human: a three-factor theory of anthropomorphism.},
	volume = {114},
	number = {4},
	journal = {Psychological review},
	publisher = {American Psychological Association},
	author = {Epley, Nicholas and Waytz, Adam and Cacioppo, John T.},
	year = {2007},
	pages = {864},
}

@article{waytz_who_2010,
	title = {Who sees human? {The} stability and importance of individual differences in anthropomorphism},
	volume = {5},
	number = {3},
	journal = {Perspectives on psychological science},
	publisher = {Sage Publications Sage CA: Los Angeles, CA},
	author = {Waytz, Adam and Cacioppo, John and Epley, Nicholas},
	year = {2010},
	pages = {219--232},
}

@article{koban_it_2024,
	title = {It feels, therefore it is: {Associations} between mind perception and mind ascription for social robots},
	volume = {153},
	journal = {Computers in Human Behavior},
	publisher = {Elsevier},
	author = {Koban, Kevin and Banks, Jaime},
	year = {2024},
	pages = {108098},
}

@article{pentina_exploring_2023,
	title = {Exploring relationship development with social chatbots: {A} mixed-method study of replika},
	volume = {140},
	journal = {Computers in Human Behavior},
	publisher = {Elsevier},
	author = {Pentina, Iryna and Hancock, Tyler and Xie, Tianling},
	year = {2023},
	pages = {107600},
}

@article{xie_attachment_2022,
	title = {Attachment theory as a framework to understand relationships with social chatbots: a case study of {Replika}},
	author = {Xie, Tianling and Pentina, Iryna},
	year = {2022},
}

@article{laestadius_too_2024,
	title = {Too human and not human enough: {A} grounded theory analysis of mental health harms from emotional dependence on the social chatbot {Replika}},
	volume = {26},
	number = {10},
	journal = {New Media \& Society},
	publisher = {SAGE Publications Sage UK: London, England},
	author = {Laestadius, Linnea and Bishop, Andrea and Gonzalez, Michael and Illenčík, Diana and Campos-Castillo, Celeste},
	year = {2024},
	pages = {5923--5941},
}

@article{fang_how_2025,
	title = {How ai and human behaviors shape psychosocial effects of extended chatbot use: {A} longitudinal randomized controlled study},
	journal = {arXiv preprint arXiv:2503.17473},
	author = {Fang, Cathy Mengying and Liu, Auren R. and Danry, Valdemar and Lee, Eunhae and Chan, Samantha WT and Pataranutaporn, Pat and Maes, Pattie and Phang, Jason and Lampe, Michael and Ahmad, Lama},
	year = {2025},
}

@article{wienrich_long-term_2023,
	title = {Long-term effects of perceived friendship with intelligent voice assistants on usage behavior, user experience, and social perceptions},
	volume = {12},
	number = {4},
	journal = {Computers},
	publisher = {MDPI},
	author = {Wienrich, Carolin and Carolus, Astrid and Markus, André and Augustin, Yannik and Pfister, Jan and Hotho, Andreas},
	year = {2023},
	pages = {77},
}

@inproceedings{muller_chatbot_2019,
	title = {Chatbot acceptance: {A} latent profile analysis on individuals' trust in conversational agents},
	booktitle = {Proceedings of the 2019 on computers and people research conference},
	author = {Müller, Lea and Mattke, Jens and Maier, Christian and Weitzel, Tim and Graser, Heinrich},
	year = {2019},
	pages = {35--42},
}

@article{smrke_aging_2022,
	title = {Aging adults’ motivation to use embodied conversational agents in instrumental activities of daily living: results of latent profile analysis},
	volume = {19},
	number = {4},
	journal = {International Journal of Environmental Research and Public Health},
	publisher = {MDPI},
	author = {Smrke, Urška and Plohl, Nejc and Mlakar, Izidor},
	year = {2022},
	pages = {2373},
}

@article{yi_potential_2024,
	title = {Potential user segmentation based on expectations of social robots using {Q}-methodology},
	volume = {12},
	journal = {IEEE Access},
	publisher = {IEEE},
	author = {Yi, Eunju and Park, Do-Hyung},
	year = {2024},
	pages = {100295--100315},
}

@article{toivonen_varieties_2024,
	title = {The varieties of agency in human–smart device relationships: {The} four agency profiles},
	volume = {16},
	number = {3},
	journal = {Future Internet},
	publisher = {MDPI},
	author = {Toivonen, Heidi and Lelli, Francesco},
	year = {2024},
	pages = {90},
}

@article{merrill_jr_ai_2022,
	title = {{AI} companions for lonely individuals and the role of social presence},
	volume = {39},
	number = {2},
	journal = {Communication Research Reports},
	publisher = {Taylor \& Francis},
	author = {Merrill Jr, Kelly and Kim, Jihyun and Collins, Chad},
	year = {2022},
	pages = {93--103},
}

@incollection{hacker_chatbots_2025,
	edition = {1},
	title = {Chatbots as {Social} {Companions}: {How} {People} {Perceive} {Consciousness}, {Human} {Likeness}, and {Social} {Health} {Benefits} in {Machines}},
	isbn = {978-0-19-894521-5},
	shorttitle = {Chatbots as {Social} {Companions}},
	url = {https://academic.oup.com/edited-volume/59762/chapter/508604431},
	doi = {10.1093/9780198945215.003.0011},
	language = {en},
	urldate = {2026-05-18},
	booktitle = {Oxford {Intersections}: {AI} in {Society}},
	publisher = {Oxford University PressOxford},
	author = {Guingrich, Rose E and Graziano, Michael S A},
	editor = {Hacker, Philipp},
	month = mar,
	year = {2025},
}

@article{carolus_smartphones_2019,
	title = {Smartphones as digital companions: {Characterizing} the relationship between users and their phones},
	volume = {21},
	issn = {1461-4448, 1461-7315},
	shorttitle = {Smartphones as digital companions},
	url = {https://journals.sagepub.com/doi/10.1177/1461444818817074},
	doi = {10.1177/1461444818817074},
	language = {en},
	number = {4},
	urldate = {2026-05-18},
	journal = {New Media \& Society},
	author = {Carolus, Astrid and Binder, Jens F and Muench, Ricardo and Schmidt, Catharina and Schneider, Florian and Buglass, Sarah L},
	month = apr,
	year = {2019},
	pages = {914--938},
}

@article{ki_can_2020,
	title = {Can an intelligent personal assistant ({IPA}) be your friend? {Para}-friendship development mechanism between {IPAs} and their users},
	volume = {111},
	issn = {07475632},
	shorttitle = {Can an intelligent personal assistant ({IPA}) be your friend?},
	url = {https://linkinghub.elsevier.com/retrieve/pii/S0747563220301655},
	doi = {10.1016/j.chb.2020.106412},
	language = {en},
	urldate = {2026-05-18},
	journal = {Computers in Human Behavior},
	author = {Ki, Chung-Wha (Chloe) and Cho, Erin and Lee, Jung-Eun},
	month = oct,
	year = {2020},
	pages = {106412},
}

@article{de_freitas_ai_2026,
	title = {{AI} {Companions} {Reduce} {Loneliness}},
	volume = {52},
	copyright = {https://academic.oup.com/pages/standard-publication-reuse-rights},
	issn = {0093-5301, 1537-5277},
	url = {https://academic.oup.com/jcr/article/52/6/1126/8173802},
	doi = {10.1093/jcr/ucaf040},
	language = {en},
	number = {6},
	urldate = {2026-05-18},
	journal = {Journal of Consumer Research},
	author = {De Freitas, Julian and Oğuz-Uğuralp, Zeliha and Uğuralp, Ahmet Kaan and Puntoni, Stefano},
	editor = {Stephen, Andrew T and Urminsky, Oleg and Novak, Thomas P},
	month = apr,
	year = {2026},
	pages = {1126--1148},
}

@misc{liu_chatbot_2024,
	title = {Chatbot {Companionship}: {A} {Mixed}-{Methods} {Study} of {Companion} {Chatbot} {Usage} {Patterns} and {Their} {Relationship} to {Loneliness} in {Active} {Users}},
	copyright = {Creative Commons Attribution 4.0 International},
	shorttitle = {Chatbot {Companionship}},
	url = {https://arxiv.org/abs/2410.21596},
	doi = {10.48550/ARXIV.2410.21596},
	urldate = {2026-05-18},
	publisher = {arXiv},
	author = {Liu, Auren R. and Pataranutaporn, Pat and Maes, Pattie},
	year = {2024},
	note = {Version Number: 3},
}

@misc{zhang_rise_2025,
	title = {The {Rise} of {AI} {Companions}: {Interaction} with {AI} {Companions} and {Psychological} {Well}-being},
	copyright = {arXiv.org perpetual, non-exclusive license},
	shorttitle = {The {Rise} of {AI} {Companions}},
	url = {https://arxiv.org/abs/2506.12605},
	doi = {10.48550/ARXIV.2506.12605},
	urldate = {2026-05-18},
	publisher = {arXiv},
	author = {Zhang, Yutong and Zhao, Dora and Hancock, Jeffrey T. and Kraut, Robert and Yang, Diyi},
	year = {2025},
	note = {Version Number: 5},
}

@misc{ciriello_not_2026,
	title = {Not a {Silver} {Bullet} for {Loneliness}: {How} {Attachment} and {Age} {Shape} {Intimacy} with {AI} {Companions}},
	copyright = {Creative Commons Attribution Non Commercial No Derivatives 4.0 International},
	shorttitle = {Not a {Silver} {Bullet} for {Loneliness}},
	url = {https://arxiv.org/abs/2602.12476},
	doi = {10.48550/ARXIV.2602.12476},
	urldate = {2026-05-18},
	publisher = {arXiv},
	author = {Ciriello, Raffaele and Gal, Uri and Turel, Ofir},
	year = {2026},
	note = {Version Number: 1},
}

@article{janssen_psychological_2024,
	title = {A {Psychological} {Need}-{Fulfillment} {Perspective} for {Designing} {Social} {Robots} that {Support} {Well}-{Being}},
	volume = {16},
	issn = {1875-4791, 1875-4805},
	url = {https://link.springer.com/10.1007/s12369-024-01102-8},
	doi = {10.1007/s12369-024-01102-8},
	language = {en},
	number = {5},
	urldate = {2026-05-18},
	journal = {International Journal of Social Robotics},
	author = {Janssen, Suzanne and Schadenberg, Bob R.},
	month = may,
	year = {2024},
	pages = {857--878},
}

@article{moradbakhti_development_2024,
	title = {Development and {Validation} of a {Basic} {Psychological} {Needs} {Scale} for {Technology} {Use}},
	volume = {5},
	copyright = {https://creativecommons.org/licenses/by/4.0},
	issn = {2698-1866},
	url = {https://econtent.hogrefe.com/doi/10.1027/2698-1866/a000062},
	doi = {10.1027/2698-1866/a000062},
	language = {en},
	number = {1},
	urldate = {2026-05-18},
	journal = {Psychological Test Adaptation and Development},
	author = {Moradbakhti, Laura and Leichtmann, Benedikt and Mara, Martina},
	month = jan,
	year = {2024},
	pages = {26--45},
}

@article{machia_social_2024,
	title = {Social need fulfillment model for human–{AI} relationships.},
	volume = {5},
	issn = {2689-0208},
	url = {https://doi.apa.org/doi/10.1037/tmb0000141},
	doi = {10.1037/tmb0000141},
	language = {en},
	number = {4},
	urldate = {2026-05-18},
	journal = {Technology, Mind, and Behavior},
	author = {Machia, Laura V. and Corral, Daniel and Jakubiak, Brett K.},
	year = {2024},
	pages = {231--239},
}

@article{gray_dimensions_2007,
	title = {Dimensions of {Mind} {Perception}},
	volume = {315},
	issn = {0036-8075, 1095-9203},
	url = {https://www.science.org/doi/10.1126/science.1134475},
	doi = {10.1126/science.1134475},
	language = {en},
	number = {5812},
	urldate = {2026-05-18},
	journal = {Science},
	author = {Gray, Heather M. and Gray, Kurt and Wegner, Daniel M.},
	month = feb,
	year = {2007},
	pages = {619--619},
}

@article{aron_inclusion_1992,
	title = {Inclusion of {Other} in the {Self} {Scale} and the structure of interpersonal closeness.},
	volume = {63},
	issn = {1939-1315, 0022-3514},
	url = {https://doi.apa.org/doi/10.1037/0022-3514.63.4.596},
	doi = {10.1037/0022-3514.63.4.596},
	language = {en},
	number = {4},
	urldate = {2026-05-18},
	journal = {Journal of Personality and Social Psychology},
	author = {Aron, Arthur and Aron, Elaine N. and Smollan, Danny},
	month = oct,
	year = {1992},
	pages = {596--612},
}

@article{voss_measuring_2003,
	title = {Measuring the {Hedonic} and {Utilitarian} {Dimensions} of {Consumer} {Attitude}},
	volume = {40},
	copyright = {https://journals.sagepub.com/page/policies/text-and-data-mining-license},
	issn = {0022-2437, 1547-7193},
	url = {https://journals.sagepub.com/doi/10.1509/jmkr.40.3.310.19238},
	doi = {10.1509/jmkr.40.3.310.19238},
	language = {en},
	number = {3},
	urldate = {2026-05-18},
	journal = {Journal of Marketing Research},
	author = {Voss, Kevin E. and Spangenberg, Eric R. and Grohmann, Bianca},
	month = aug,
	year = {2003},
	pages = {310--320},
}

@article{chen_basic_2015,
	title = {Basic psychological need satisfaction, need frustration, and need strength across four cultures},
	volume = {39},
	issn = {0146-7239, 1573-6644},
	url = {http://link.springer.com/10.1007/s11031-014-9450-1},
	doi = {10.1007/s11031-014-9450-1},
	language = {en},
	number = {2},
	urldate = {2026-05-18},
	journal = {Motivation and Emotion},
	author = {Chen, Beiwen and Vansteenkiste, Maarten and Beyers, Wim and Boone, Liesbet and Deci, Edward L. and Van Der Kaap-Deeder, Jolene and Duriez, Bart and Lens, Willy and Matos, Lennia and Mouratidis, Athanasios and Ryan, Richard M. and Sheldon, Kennon M. and Soenens, Bart and Van Petegem, Stijn and Verstuyf, Joke},
	month = apr,
	year = {2015},
	pages = {216--236},
}

@article{hughes_short_2004,
	title = {A {Short} {Scale} for {Measuring} {Loneliness} in {Large} {Surveys}: {Results} {From} {Two} {Population}-{Based} {Studies}},
	volume = {26},
	copyright = {https://journals.sagepub.com/page/policies/text-and-data-mining-license},
	issn = {0164-0275, 1552-7573},
	shorttitle = {A {Short} {Scale} for {Measuring} {Loneliness} in {Large} {Surveys}},
	url = {https://journals.sagepub.com/doi/10.1177/0164027504268574},
	doi = {10.1177/0164027504268574},
	language = {en},
	number = {6},
	urldate = {2026-05-18},
	journal = {Research on Aging},
	author = {Hughes, Mary Elizabeth and Waite, Linda J. and Hawkley, Louise C. and Cacioppo, John T.},
	month = nov,
	year = {2004},
	pages = {655--672},
}

@article{baumeister_need_1995,
	title = {The need to belong: {Desire} for interpersonal attachments as a fundamental human motivation.},
	volume = {117},
	issn = {1939-1455, 0033-2909},
	shorttitle = {The need to belong},
	url = {https://doi.apa.org/doi/10.1037/0033-2909.117.3.497},
	doi = {10.1037/0033-2909.117.3.497},
	language = {en},
	number = {3},
	urldate = {2026-05-18},
	journal = {Psychological Bulletin},
	author = {Baumeister, Roy F. and Leary, Mark R.},
	month = may,
	year = {1995},
	pages = {497--529},
}

@article{laursen_person-centered_2006,
	title = {Person-centered and variable-centered approaches to longitudinal data},
	volume = {52},
	number = {3},
	journal = {Merrill-Palmer Quarterly},
	publisher = {Wayne State University Press},
	author = {Laursen, Brett Paul and Hoff, Erika},
	year = {2006},
	pages = {377--389},
}

@article{bergman_person-oriented_1997,
	title = {A person-oriented approach in research on developmental psychopathology},
	volume = {9},
	number = {2},
	journal = {Development and psychopathology},
	publisher = {Cambridge University Press},
	author = {Bergman, Lars R. and Magnusson, David},
	year = {1997},
	pages = {291--319},
}

@article{spurk_latent_2020,
	title = {Latent profile analysis: {A} review and “how to” guide of its application within vocational behavior research},
	volume = {120},
	journal = {Journal of vocational behavior},
	publisher = {Elsevier},
	author = {Spurk, Daniel and Hirschi, Andreas and Wang, Mo and Valero, Domingo and Kauffeld, Simone},
	year = {2020},
	pages = {103445},
}

@article{weizenbaum_elizacomputer_1966,
	title = {{ELIZA}—a computer program for the study of natural language communication between man and machine},
	volume = {9},
	number = {1},
	journal = {Communications of the ACM},
	publisher = {ACM New York, NY, USA},
	author = {Weizenbaum, Joseph},
	year = {1966},
	pages = {36--45},
}

@article{weizenbaum_computer_1976,
	title = {Computer power and human reason: {From} judgment to calculation},
	journal = {San Francisco},
	author = {Weizenbaum, Joseph},
	year = {1976},
}

@article{reeves_media_1996,
	title = {The media equation: {How} people treat computers, television, and new media like real people},
	volume = {10},
	number = {10},
	journal = {Cambridge, UK},
	author = {Reeves, Byron and Nass, Clifford},
	year = {1996},
	pages = {19--36},
}

@article{van_der_goot_disentangling_2023,
	title = {Disentangling two fundamental paradigms in human-machine communication research: {Media} equation and media evocation},
	volume = {6},
	number = {1},
	journal = {Human-Machine Communication},
	author = {van der Goot, Margot J. and Etzrodt, Katrin},
	year = {2023},
	pages = {2},
}

@article{banks_toward_2020,
	title = {Toward an {Agent}-{Agnostic} {Transmission} {Model}: {Synthesizing} {Anthropocentric} and {Technocentric} {Paradigms} in {Communication}},
	volume = {1},
	issn = {2638-6038, 2638-602X},
	shorttitle = {Toward an {Agent}-{Agnostic} {Transmission} {Model}},
	url = {https://stars.library.ucf.edu/hmc/vol1/iss1/2/},
	doi = {10.30658/hmc.1.2},
	language = {en},
	urldate = {2026-08-28},
	journal = {Human-Machine Communication},
	author = {Banks, Jamie and De Graaf, Maartje},
	month = feb,
	year = {2020},
	pages = {19--36},
}

@inproceedings{banks_ghosting_2026,
	address = {Bremen Germany},
	title = {Ghosting the {Machine}: {Stop} {Calling} {Human}-{Agent} {Relations} {Parasocial}},
	isbn = {979-8-4007-2741-2},
	shorttitle = {Ghosting the {Machine}},
	url = {https://dl.acm.org/doi/10.1145/3816046.3816308},
	doi = {10.1145/3816046.3816308},
	language = {en},
	urldate = {2026-08-28},
	booktitle = {Proceedings of the 8th {ACM} {Conference} on {Conversational} {User} {Interfaces}},
	publisher = {ACM},
	author = {Banks, Jaime},
	month = jul,
	year = {2026},
	pages = {1--6},
}

@incollection{norskov_sociomorphing_2020,
	title = {Sociomorphing, {Not} {Anthropomorphizing}: {Towards} a {Typology} of {Experienced} {Sociality}},
	isbn = {978-1-64368-154-2 978-1-64368-155-9},
	shorttitle = {Sociomorphing, {Not} {Anthropomorphizing}},
	url = {http://ebooks.iospress.nl/doi/10.3233/FAIA200900},
	doi = {10.3233/FAIA200900},
	urldate = {2026-08-28},
	booktitle = {Frontiers in {Artificial} {Intelligence} and {Applications}},
	publisher = {IOS Press},
	author = {Seibt, Johanna and Vestergaard, Christina and Damholdt, Malene F.},
	editor = {Nørskov, Marco and Seibt, Johanna and Quick, Oliver Santiago},
	month = dec,
	year = {2020},
}

@book{bowlby_attachment_1969,
	title = {Attachment and loss vol. 3},
	author = {Bowlby, John},
	year = {1969},
}

@article{ainsworth_patterns_1978,
	title = {Patterns of},
	journal = {Attachment: A},
	author = {Ainsworth, M. D. S. and Blehar, M. C. and Waters, E. and Wall, S.},
	year = {1978},
}

@article{hazan_romantic_1987,
	title = {Romantic love conceptualized as an attachment process.},
	volume = {52},
	number = {3},
	journal = {Journal of personality and social psychology},
	publisher = {American Psychological Association},
	author = {Hazan, Cindy and Shaver, Phillip},
	year = {1987},
	pages = {511},
}

@book{mikulincer_attachment_2010,
	title = {Attachment in adulthood: {Structure}, dynamics, and change},
	isbn = {1-60623-610-5},
	publisher = {Guilford Publications},
	author = {Mikulincer, Mario and Shaver, Phillip R.},
	year = {2010},
}

@article{gillath_attachment_2021,
	title = {Attachment and trust in artificial intelligence},
	volume = {115},
	journal = {Computers in human behavior},
	publisher = {Elsevier},
	author = {Gillath, Omri and Ai, Ting and Branicky, Michael S. and Keshmiri, Shawn and Davison, Robert B. and Spaulding, Ryan},
	year = {2021},
	pages = {106607},
}

@misc{banks_measuring_2025,
	type = {Data set},
	title = {Measuring {Machine} {Companionship} {Experiences}: {Scale} {Development} and {Initial} {Validation} for {AI} {Companion}},
	url = {https://osf.io/hjg2a/},
	publisher = {OSF},
	author = {Banks, Jaime},
	month = oct,
	year = {2025},
}

@article{vermunt_lg-syntax_2021,
	title = {{LG}-syntax user’s guide: manual for latent gold syntax module version 6.0},
	journal = {Arlington, MA: Statistical Innovations Inc},
	author = {Vermunt, JEROEN K. and Magidson, J.},
	year = {2021},
}

@article{nylund_deciding_2007,
	title = {Deciding on the number of classes in latent class analysis and growth mixture modeling: {A} {Monte} {Carlo} simulation study},
	volume = {14},
	number = {4},
	journal = {Structural equation modeling: A multidisciplinary Journal},
	publisher = {Taylor \& Francis},
	author = {Nylund, Karen L. and Asparouhov, Tihomir and Muthén, Bengt O.},
	year = {2007},
	pages = {535--569},
}

@article{bakk_robustness_2016,
	title = {Robustness of {Stepwise} {Latent} {Class} {Modeling} {With} {Continuous} {Distal} {Outcomes}},
	volume = {23},
	issn = {1070-5511, 1532-8007},
	url = {http://www.tandfonline.com/doi/full/10.1080/10705511.2014.955104},
	doi = {10.1080/10705511.2014.955104},
	language = {en},
	number = {1},
	urldate = {2026-08-28},
	journal = {Structural Equation Modeling: A Multidisciplinary Journal},
	author = {Bakk, Zsuzsa and Vermunt, Jeroen K.},
	month = jan,
	year = {2016},
	pages = {20--31},
}

@article{vermunt_latent_2010,
	title = {Latent {Class} {Modeling} with {Covariates}: {Two} {Improved} {Three}-{Step} {Approaches}},
	volume = {18},
	copyright = {https://www.cambridge.org/core/terms},
	issn = {1047-1987, 1476-4989},
	shorttitle = {Latent {Class} {Modeling} with {Covariates}},
	url = {https://www.cambridge.org/core/product/identifier/S1047198700012560/type/journal_article},
	doi = {10.1093/pan/mpq025},
	language = {en},
	number = {4},
	urldate = {2026-08-28},
	journal = {Political Analysis},
	author = {Vermunt, Jeroen K.},
	year = {2010},
	pages = {450--469},
}

@article{ebner_understanding_2026,
	title = {Understanding romantic relationships between humans and chatbots: {A} qualitative and quantitative study on romantic fantasy and other interpersonal characteristics.},
	volume = {7},
	issn = {2689-0208},
	shorttitle = {Understanding romantic relationships between humans and chatbots},
	url = {https://doi.apa.org/doi/10.1037/tmb0000193},
	doi = {10.1037/tmb0000193},
	language = {en},
	number = {2},
	urldate = {2026-09-06},
	journal = {Technology, Mind, and Behavior},
	author = {Ebner, Paula and Szczuka, Jessica},
	month = jun,
	year = {2026},
	pages = {83--97},
}

@article{aron_self-expansion_2022,
	title = {Self-expansion motivation and inclusion of others in self: {An} updated review},
	volume = {39},
	issn = {0265-4075, 1460-3608},
	shorttitle = {Self-expansion motivation and inclusion of others in self},
	url = {https://journals.sagepub.com/doi/10.1177/02654075221110630},
	doi = {10.1177/02654075221110630},
	language = {en},
	number = {12},
	urldate = {2026-09-06},
	journal = {Journal of Social and Personal Relationships},
	author = {Aron, Arthur and Lewandowski, Gary and Branand, Brittany and Mashek, Debra and Aron, Elaine},
	month = dec,
	year = {2022},
	pages = {3821--3852},
}

@inproceedings{manoli_2026_digital,
  title={Digital Companionship: Overlapping Uses of {AI} Companions and {AI} Assistants},
  author={Manoli, Aikaterina and Pauketat, Janet V. T. and Ladak, Ali and Noh, Hayoun and Hwang, Angel Hsing-Chi and Anthis, Jacy Reese},
  booktitle={Proceedings of the 2026 CHI Conference on Human Factors in Computing Systems},
  series={CHI '26},
  year={2026},
  publisher={Association for Computing Machinery},
  address={New York, NY, USA},
  numpages={25},
  doi={10.1145/3772318.3791331}
}

@inproceedings{clark_2019_what,
  title={What Makes a Good Conversation? Challenges in Designing Truly Conversational Agents},
  author={Clark, Leigh and Pantidi, Nadia and Cooney, Orla and Doyle, Philip and Garaialde, Diego and Edwards, Justin and Spillane, Brendan and Gilmartin, Emer and Murad, Christine and Munteanu, Cosmin and Wade, Vincent and Cowan, Benjamin R.},
  booktitle={Proceedings of the 2019 CHI Conference on Human Factors in Computing Systems},
  series={CHI '19},
  articleno={475},
  numpages={12},
  year={2019},
  publisher={Association for Computing Machinery},
  address={New York, NY, USA},
  doi={10.1145/3290605.3300705}
}

@article{schimmelpfennig_2026_humanlike,
  title={Humanlike {AI} Design Increases Anthropomorphism but Yields Divergent Outcomes on Engagement and Trust Globally},
  author={Schimmelpfennig, Robin and Diaz, Mark and Prabhakaran, Vinodkumar and Davani, Aida},
  journal={arXiv preprint arXiv:2512.17898},
  year={2026},
  doi={10.48550/arXiv.2512.17898}
}

\end{document}